\documentstyle[epsf,12pt]{article}
\newcommand{\de}{\partial}
\newcommand{\be}{\begin{equation}}
\newcommand{\ba}{\begin{eqnarray}}
\newcommand{\ea}{\end{eqnarray}}
\newcommand{\ee}{\end{equation}}

\newcommand{\we}{\wedge}
\newcommand{\ca}{\mathcal}

\newcommand{\f}{\frac}
\newcommand{\s}{\sqrt}

\newcommand{\ap}{\alpha}

\newcommand{\mb}{\mathbf}
\newcommand{\ddd}{\cdot\cdot\cdot}
\newcommand{\no}{\nonumber \\}

\newcommand{\la}{\langle}
\newcommand{\lb}{\rangle}
\newcommand{\ep}{\epsilon}

\newcommand{\ov}{\overline}

\begin{document}
\begin{titlepage}
\thispagestyle{empty}
\begin{flushright}
hep-th/0106142 \\
UT-946 \\
June, 2001 \\
\end{flushright}

\bigskip
\bigskip

\begin{center}
\noindent{\Large \textbf{Tachyon Condensation on Orbifolds and}}\\
\noindent{\Large \textbf{McKay Correspondence}}\\
\bigskip
\bigskip
\noindent{
          Tadashi Takayanagi\footnote{
                 E-mail: takayana@hep-th.phys.s.u-tokyo.ac.jp} }\\

\bigskip
{Department of Physics, Faculty of Science \\ University of Tokyo \\
\medskip
Bunkyo-ku, Tokyo 113-0033, Japan}

\end{center}
\begin{abstract}
In this short note we present a mathematical interpretation of
 tachyon condensation on (three dimensional) orbifolds within the
framework of boundary string field theory (BSFT). 
We explicitly show that important parts of decay modes 
in brane-antibrane systems with ${\ca N}=2$ boundary supersymmetry
 can be interpreted as the McKay correspondence described as complexes.
This will give an example of
 the recent interpretation of D-branes as derived category. 
We also discuss the ${\ca N}=4$ boundary supersymmetry as a more 
refined structure.
\end{abstract}
\end{titlepage}

\newpage

\section{Introduction}
\setcounter{equation}{0}
\hspace{5mm}
The studies of tachyon condensation in open string theory \cite{Sen985}
have revealed important aspects
of D-branes\footnote{For earlier work on tachyon condensation see 
\cite{Hal}.}. One of these is the classification of D-brane
charges as $K$-theory \cite{Witten3}. Brane-antibrane systems can be 
considered topologically as the difference of two bundles and this fact just
corresponds to the mathematical definition of $K$-theory. 
As a next step 
it is natural to ask whether we can go beyond this topological argument
by assuming more refined structures such as world-sheet extended 
supersymmetries. Recently the interpretation of a D-brane on complex 
manifolds
 as a complex of coherent sheaves has been discussed (for example see 
 \cite{Sh,Oz,Do,As}). A coherent sheaf ${\ca E}$ is a more general and 
 refined
 notion than 
 a vector
 bundle (locally free sheaf). For example, it admits `almost rank=0' or 
 point-like instanton configurations. It includes its cokernel ${\ca E}''$ 
 defined by the exact sequence $0\to{\ca E}\to{\ca E}'\to{\ca E}''\to 0$ and 
 this can be regarded as tachyon condensation on a brane ${\ca E}$ and an 
 antibrane 
 ${\ca E}'$ into the smaller brane ${\ca E}''$.
These considerations eventually lead to the 
interpretation of D-branes as derived category \cite{Do,As,Sh}.
 
In this paper we would like to discuss such an interpretation for
tachyon condensation on brane-antibrane systems in orbifold theories. 
In particular we describe the tachyon condensation in the framework of 
boundary string
field theory (BSFT) \cite{bsft,GeSh,KuMaMo2,KrLa,TaTeUe}. This theory is 
very useful because it is believed that infinitely many massive fields 
are not relevant for tachyon condensation. Furthermore we require the 
boundary
${\ca N}=2$ (B-type) supersymmetry \cite{OoOzYi,Ho} so as to study the
 algebra geometric aspects of D-branes. The general strategy of applying BSFT
 to orbifold theories has already been discussed in the
  previous paper \cite{holtak}. Here we investigate especially
 brane-antibrane systems on three dimensional orbifolds in detail. 
{}From this analysis we will find that 
the tachyon condensation can be analyzed in a group
 theoretical way and the essential decay processes are described by 
 the complexes which describe the McKay correspondence 
 \cite{ItNa} proved by Ito and Nakajima 
 (see \cite{Mc} for the original paper and also \cite{Re} 
 for a review). We will also discuss the ${\ca N}=4$ boundary supersymmetry
 as a more refined structure and see that this symmetry 
 leads to a certain quaternionic
constraint on tachyon fields.

\section{Tachyon Condensation on Orbifolds as McKay Correspondence}
\hspace{5mm}
\setcounter{equation}{0}
\hspace{5mm}
Here we discuss the interpretation of tachyon condensation on orbifolds
 as the McKay correspondence \cite{ItNa,Mc,Re}. 
Though our arguments below can be easily generalized into arbitrary 
dimensional abelian orbifolds, we describe the results for 
D6-$\ov{\mbox{D6}}$ systems in
three dimensional abelian
orbifolds ${\bf C^3}/{\bf \Gamma}$. This is not only because these are 
the most physically interesting but because
the corresponding 
mathematical results on the resolution of orbifold singularities 
have not been obtained in higher dimensions. 

Let us first review the outline of the analysis in \cite{holtak}. Since the 
physics of tachyon condensation involves off-shell string theory, 
one should 
consider within the framework of string field theory. We apply the boundary 
string field theory (BSFT) \cite{bsft} for brane-antibrane systems 
\cite{KrLa,TaTeUe} to tachyon condensation on orbifolds.

We begin with the disk world-sheet action in flat background with boundary 
interactions which preserve ${\ca{N}}=2$ (B-type) boundary supersymmetry 
\cite{Ho}.
We denote the ten dimensional complex coordinates of the target space as 
$({ Z}_1,\ov{{ Z}_1})
 ,\ddd,({ Z}_5,\ov{{ Z}_5})$. 
The explicit expression of the boundary interactions is given as follows:
\ba
I_B &=& -\int_{\de\Sigma} d\tau d\theta d\bar{\theta}\sum_{i}
{\bf{\Gamma}}_i\bar{{\bf \Gamma}_i}+\int_{\de\Sigma} d\tau d\theta \frac{1}
{\sqrt{2\pi}}\sum_{i}{\bf \Gamma}_iT_i({\bf Z})+(\mbox{h.c.}), \label{bim}
\ea
where we have employed ${\ca{N}}=2$ boundary superspace 
$(\tau,\theta,\bar{\theta})$; the boundary fermionic 
chiral and antichiral superfields 
${\bf \Gamma}_i,\ov{{\bf \Gamma}}_i$ are defined in our conventions as
\ba
{\bf \Gamma}_i&=&-\f{i}{\s{2}}\eta_i+\theta F_i-
\f{i}{\s{2}}\theta\bar{\theta}
\de_{\tau}\eta_i, \no
\ov{{\bf \Gamma}}_i&=&\f{i}{\s{2}}\bar{\eta}_i+\theta \bar{F_i}-\f{i}{\s{2}}
\theta\bar{\theta}
\de_{\tau}\bar{\eta}_i.
\ea
The fermions $\eta_i,\bar{\eta}_i$ are called boundary fermions and the 
scalar fields $F_i,\bar{F_i}$ are auxiliary fields.
Since we are interested in the decay of a D6-$\ov{\mbox{D6}}$ system into 
D0-branes below, we need $2^3/2=4$ pairs of D6-$\ov{\mbox{D6}}$. 
We can express this 
brane-antibrane system by using three fermi superfields ${\bf \Gamma}_i \ \ 
(i=1,2,3)$.

Note that the world-sheet ${\ca{N}}=2$ supersymmetry requires the form of
tachyon fields $T_i({\bf Z})$ to be holomorphic and this means that we 
consider
 a certain subspace of the field space of BSFT. This restriction is essential
 in our arguments not only because one can apply non-renomalization 
 theorem but because it matches with complex geometry.

The spacetime action $S$ of BSFT for brane-antibrane system is defined to 
be equal to the disk partition function $Z$ \cite{Ts,KuMaMo2}. 
For example, if one 
considers the following tachyon fields\footnote{
If one considers more general holomorphic configurations than eq.(\ref{tak}),
then some brane-antibrane systems will be generated after the tachyon 
condensation as observed in \cite{holtak}.}
on the D6-$\ov{\mbox{D6}}$ system
\ba
T_1=(Z_1)^p,\ \ T_2=(Z_2)^q,\ \ T_3=(Z_3)^r, \label{tak}
\ea
then we obtain 
$pqr$ D0-branes after the tachyon condensation. This can be shown by 
computing
 spacetime action $S$ employing the non-renomalization 
 theorem \cite{Ho} or calculating the RR-couplings \cite{holtak} 
 expressed by Quillen's 
 superconnection 
 \cite{KeWi,KrLa,TaTeUe}. In particular the tachyon field $p=q=r=1$ is
 known as Atiyah-Bott-Shapiro configuration \cite{ABS}, 
 which carries the unit $K$-theory
 charge. This configuration will also play an very 
 important role in the orbifold
 theories discussed below.

The above arguments of tachyon condensation on flat spaces can be generalized
 to orbifold theories.
In the previous paper \cite{holtak}, we defined the BSFT for brane-antibrane
systems on (abelian) orbifolds following the general framework of 
\cite{DoMo} and 
analyzed the examples in two 
dimensional orbifolds ${\bf C^2}/{\bf Z}_N$ in detail. From the results in 
BSFT
one can determine how many fractional D0-branes \cite{DiDoGo} 
are generated after the 
tachyon
condensation, but it is difficult to see 
what kinds of fractional D0-branes they are. Therefore we were also needed to 
apply the conservation law of twisted RR-charges.

Now let us discuss the tachyon condensation on a three dimensional orbifold
${\bf C^3}/{\bf \Gamma}$. We assume that the discrete group ${\bf \Gamma}$ 
is abelian
 because it is difficult to describe non-abelian orbifold actions in the 
 boundary interaction (\ref{bim}). For example, if ${\bf \Gamma}={\bf Z}_N$,
  then the action on the coordinate $(z^1,z^2,z^3)$ of ${\bf C}^3$ is defined
 as follows.
\ba
g: (z^1,z^2,z^3) \to (e^{\f{2\pi i a_1}{N}}z^1,e^{\f{2\pi i a_2}{N}}z^2,
e^{\f{2\pi i a_3}{N}}z^3),
\ea
where $(a_1,a_2,a_3)$ are integers which 
satisfy $a_1+a_2+a_3=0 \ (\mbox{mod}N)$. We define
such a three dimensional fundamental representation as $Q$.
Generally, D-branes in orbifold theories are classified by 
the representations of the group ${\bf \Gamma}$. Here we are interested in
fractional D-branes and these correspond to the irreducible 
representations. For the orbifold ${\bf \Gamma}={\bf Z}_N$, they are 
given by
 one dimensional representations 
$\{\rho_{\ap}\}\ \ (\ap=0,1\ddd,N-1)$, which are defined by the 
multiplication of
the phase factor $e^{\f{2\pi i \ap}{N}}$. 
Therefore we will express each type of
 these fractional D-branes by $\rho_{\ap}$ below .
Though the following discussion can be applied to any abelian 
orbifold action, we mainly consider ${\bf \Gamma}={\bf Z}_N$ case.

Then let us
 examine the orbifold action on boundary fields. For this purpose it is 
 useful 
 to rewrite the interaction (\ref{bim}) into non-abelian tachyon fields $T$
by employing the interpretation of ${\bf \Gamma}_i,{\bf \ov{\Gamma}}_i$ as 
  $\gamma$-matrices\footnote{We have defined the matrices 
  $\gamma^{+}_{i},\gamma^{-}_{i}$ such that $\{\gamma^{+}_{i},
  \gamma^{-}_{j}\}
  =\delta_{ij},\ \{\gamma^{+}_{i},\gamma^{+}_{j}\}=
  \{\gamma^{-}_{i},\gamma^{-}_{j}\}=0$.}
 $\gamma^{+}_{i},\gamma^{-}_{i}$:
\ba
T&=&\sum_{i=1}^{3}(\gamma^{+}_{i}T_i+\gamma^{-}_{i}\ov{T_i})=
\left[
	\begin{array}{cccc}
	T_3 & 0 & T_2 & T_1  \\
	0  & T_3 & \ov{T_1} & -\ov{T_2} \\
	\ov{T_2} & T_1 & -\ov{T_3} & 0 \\
	\ov{T_1} & -T_2 & 0 & -\ov{T_3}  
\end{array}
\right] \no
&=&\left[
	\begin{array}{cccc}
	(Z_3)^r & 0 & (Z_2)^q & (Z_1)^p  \\
	0  & (Z_3)^r & (\ov{Z_1})^p & -(\ov{Z_2})^q \\
	(\ov{Z_2})^q & (Z_1)^p & -(\ov{Z_3})^r & 0 \\
	(\ov{Z_1})^p & -(Z_2)^q & 0 & -(\ov{Z_3})^r  \label{tm}
\end{array}
\right].
\ea

{}From this we can consider the orbifold action on the 
Chan-Paton factors \cite{DoMo}
 for the tachyon fields (\ref{tak}). The invariance under
  this action requires us that the 
  original D6-$\ov{\mbox{D6}}$ system should consist of four D6-branes 
  $\rho_{\ap},\rho_{\ap+pa_1+qa_2},\rho_{\ap+qa_2+ra_3},
  \rho_{\ap+ra_3+pa_1}$ 
  (corresponding to each row of the matrix (\ref{tm})) and
  four anti D6-branes 
  $\rho_{\ap+ra_3},\rho_{\ap},\rho_{\ap+qa_2},\rho_{\ap+pa_1}$
  (corresponding to each column of the matrix (\ref{tm})). In this way the
  requirement of boundary ${\ca{N}}=2$ supersymmetry determines the types of
  6-branes before tachyon condensation up to an arbitrary integer $\ap$. 
  
 The 
  orbifold action on the boundary fields is given by 
\ba
g:{\bf \Gamma}_{i}\to {\bf \Gamma}_{i}e^{-\f{2\pi ia_i}{N}},
\ea
and by this action we define the BSFT for orbifolded brane-antibrane 
systems \cite{holtak}. Then we obtain $pqr$ fractional D0-branes after the
tachyon condensation. In order to determine the types of fractional D0-branes
 one needs to examine the twisted RR-charges. 
 
Here let us discuss twisted RR-charges in a group 
theoretical way. The value of these can be read \cite{DiFrPeScLeRu} 
from the
explicit forms of boundary states in twisted sectors as shown in 
\cite{orbi,holtak}. Then the RR-charges $Q^{Dp}_{\ap}(g)$ 
in $g\in {\bf \Gamma}$ twisted sector
 for a  D$p$-brane $\rho_{\ap}$ are given by 
(with an appropriate normalization)
\ba
&& Q^{D6}_{\ap,k}=\f{1}{|{\bf\Gamma}|}\chi_{\ap}(g)\s{\mbox{Tr}(g)},\no
&& Q^{D0}_{\ap,k}=\f{1}{|{\bf\Gamma}|}\chi_{\ap}(g),
\ea
where $\chi_{\ap}(g)$ denotes the character for the irreducible
 representation $\rho_{\ap}$ of ${\bf \Gamma}$.
The factor $\mbox{Tr}(g)$ comes from the trace of the bosonic 
zeromodes in open string sector\footnote{
Here we have normalized the bosonic 
zeromodes as $\la z|z'\lb=\delta^2(z-z')$.}
and is given by
\ba
\mbox{Tr}(g)&=&\int (dz)^3(d\bar{z})^3 \la z_i|g|z_i\lb  \no
&=&\left(\det(1-g)\right)^{-2}\no
&=&(1-\chi_{Q}(g)+\chi_{Q\we Q}(g)-\chi_{Q\we Q\we Q}(g))^{-2},
\ea
where $\chi_{Q}$ denotes the character for the fundametal represetation $Q$ 
and the symbol $\we$ means wedge product of representations.
Thus we obtain the following result 
using the formula $\chi_{Q}(g)\chi_{\ap}(g)=\chi_{Q\otimes
\rho_{\ap}}(g)$
\ba
Q^{D0}_{\ap}(g)&=&(1-\chi_{Q}(g)+\chi_{Q\we Q}(g)-\chi_{Q\we Q\we Q}(g))
Q^{D6}_{\ap}(g)\no
&=&\sum_{\beta}(a^{(0)}_{\beta\ap}-a^{(1)}_{\beta\ap}+a^{(2)}_{\beta\ap}
-a^{(3)}_{\beta\ap})Q^{D6}_{\beta}(g), \label{rr}
\ea
where the integers $a^{(i)}_{\beta\ap}$ are defined by the following
decomposition of representations
\ba
\we^{i}Q\otimes \rho_{\ap}=\sum_{\beta}a^{(i)}_{\beta\ap}\ \rho_{\beta}.
\ea
This result (\ref{rr}) shows that the twisted RR charges 
(or $K$-theory charges) of the fractional D0-brane $\rho_{\ap}$ 
are equal to that of the
brane-antibrane system which consists of 
$\sum_{\beta}(a^{(0)}_{\beta\ap}+a^{(2)}_{\beta\ap})\rho_{\beta}$  
D6-branes and 
$\sum_{\beta}(a^{(1)}_{\beta\ap}+a^{(3)}_{\beta\ap})\rho_{\beta}$ 
antiD6-branes.
Note that the above results hold for any (generally non-abelian) 
discrete group 
${\bf \Gamma}\in 
\mbox{SL}(3,{\bf C})$. 

If we return the case ${\bf \Gamma}={\bf Z}_N$ again, it
is easy to see 
\ba
& &a^{(0)}_{\beta\ap}=a^{(3)}_{\beta\ap}=1,\ \ 
\sum_{\beta}a^{(1)}_{\beta\ap}\rho_{\beta}=\rho_{\ap+a_1}
\oplus\rho_{\ap+a_2}\oplus\rho_{\ap+a_3}, \no
& &\sum_{\beta}a^{(2)}_{\beta\ap}\rho_{\beta}=\rho_{\ap+a_1+a_2}
\oplus\rho_{\ap+a_2+a_3}\oplus\rho_{\ap+a_3+a_1}.\label{zn}
\ea
In particular if we consider the specific tachyon fields $p=q=r=1$ in 
eq.(\ref{tm}), then 
the above results (\ref{rr}) and (\ref{zn}) show that a fractional D0-brane 
$\rho_{\ap}$ will be generated and this result is consistent 
with the previous result from BSFT. Even though the calculation of 
twisted RR-charges seems very powerful to determine the decay product 
at first sight, it should be noted that this cannot distinguish `$0$' from
`$Q^{D6}_{\ap}(g)-Q^{D6}_{\ap}(g)$'. As we have seen, this ambiguity
is fixed by the consideration of BSFT.
For general $p=p_1,\ q=p_2,\ r=p_3$ one can show the following identity:
\ba
&&\sum_{\beta_i=0}^{p_i-1}Q^{D0}_{\ap+a_1\beta_1+a_2\beta_2+a_3\beta_3}(g)
=Q^{D6}_{\ap}(g)-
\left(Q^{D6}_{\ap+pa_1}(g)+Q^{D6}_{\ap+qa_2}(g)+
Q^{D6}_{\ap+ra_3}(g)\right)  \no 
&&\ \ \ \ \ \ \ \ \ \ \ \ \ \ \ \ \ \
+\left(Q^{D6}_{\ap+pa_1+qa_2}(g)+Q^{D6}_{\ap+qa_2+ra_3}(g)+
Q^{D6}_{\ap+ra_3+pa_1}(g)\right)-Q^{D6}_{\ap}(g) \label{gt}. 
\ea
Thus we can conclude the $pqr=p_1p_2p_3$ 
fractional D0-branes which are represented by
 left-handed side of eq.(\ref{gt}) are 
 generated after tachyon condensation on the D6-$\ov{\mbox{D6}}$ system 
 defined by the right-handed side.

In this way we have determined the decay processes of brane-antibrane systems
 on orbifolds. 
 We have also found that the tachyon condensation with 
 boundary ${\ca{N}}=2$ supersymmetry sees `holomorphy' beyond the familiar
  $K$-theory charges. As argued in \cite{Sh,Oz,Do,As} the 
  tachyon condensation 
  on brane-antibrane systems in complex manifolds
   such as Calabi-Yau manifolds can be represented by a complex of 
   coherent sheaves. More precisely, one should identify two complexes
   which have the same cohomology complex (quasi-isomorphic) 
   and we eventually obtain the notion of 
   the derived category \cite{hom} as discussed in \cite{Sh,Do,As}. 
Let us briefly review this idea. For a given (bounded) complex,
\ba
{\ca E}_n \stackrel{d_{n-1}}{\longrightarrow}{\ca E}_{n-1} \stackrel{d_{n-2}}
{\longrightarrow}{\ca E}_{n-2}\stackrel{d_{n-3}}
{\longrightarrow}\ddd \stackrel{d_{1}}
{\longrightarrow}{\ca E}_{1},
\ea
it has been proposed to interpret the coherent sheaves 
${\ca E}_{even}$ as branes 
and ${\ca E}_{odd}$ as antibranes because their cohomologies, which have 
essential information for any complex, roughly mean the `subtraction' of 
adjacent modules. The tachyon fields between the brane-antibrane system
correspond to the arrows of the complex. This can also be seen from
 the $K$-theory charge of the complex, which is given by
 $[\oplus_{n}H^{2n}]\ominus [\oplus_{n}H^{2n+1}]$. The simplest example
 of this 
 will be the Koszul complex which describes tachyon condensation on
 flat space \cite{Ho,As}. 
 Below we would like to relate 
  our BSFT description of tachyon condensation on orbifolds to the complex
 which defines McKay correspondence \cite{ItNa} as a more non-trivial 
 example. If we blow up the orbifold singularities, we can consider 
 brane-antibrane
systems in such a space as an object of the derived category of coherent 
sheaves. For 
 earlier discussions of this complex in physical context see for example
 \cite{DiDo,Go}. See also \cite{Go2} for descriptions of some
 other complexes.

The fundamental complex
 of McKay correspondence in three dimension is given by
\ba
{\ca R} \stackrel{d_3}{\longrightarrow} Q\otimes {\ca R}
\stackrel{d_2}{\longrightarrow}\we^2 Q\otimes {\ca R}
\stackrel{d_1}{\longrightarrow}\we^3 Q\otimes {\ca R}={\ca R}. \label{cp1}
\ea
The bundle ${\ca R}$ denotes $|{\bf \Gamma}|$ dimensional bundle on the 
(resolved) 
orbifold space $X\sim {\bf C^3}/{\bf \Gamma}$ 
(tautological bundle) and is defined by 
${\ca R}=P\times _{GL_G(R)}{\mb C}^{|\bf \Gamma|}$. 
The fiber ${\mb C}^{|\bf \Gamma|}$ is charged 
under the (complex) quiver gauge group $GL_G(R)$ so as to belong to the 
regular
representation.
The fibration $P$ over $X$ 
is equivalent to the quiver theory for a bulk D0-brane 
(regular
representation) \cite{DoMo} such that $P/GL_G(R)=X$.
The action $Q$ denotes the three dimensional representation given 
by
 the inclusion ${\bf \Gamma}\in \mbox{SL}(3,{\bf C})$. The boundary operator
  $d_i$ is defined by $d_i=B\we$, where $B$ denotes 
  the multiplication of the coordinate functions
  $(z_1,z_2,z_3)$. Note that it is easy to see the nilpotency 
  $d_{i}d_{i+1}=0$.
  Then the bundle ${\ca R}$ can be decomposed into $|{\bf \Gamma}|$ line 
bundles corresponding to
each irreducible representations
:
\ba
{\ca R}=\oplus_{\ap}{\ca R}_\ap\otimes \rho_{\ap},
\ea
and each of them can be naturally 
interpreted as a fractional D6-brane $\rho_{\ap}$.
Then we can divide the complex (\ref{cp1}) into $|{\bf \Gamma}|$ parts:
\ba
{\ca S}_\ap:\ -\left[\ {\ca R}_\ap \stackrel{d_3}{\longrightarrow} 
\oplus_{\beta}a^{(1)}_{\ap\beta}{\ca R}_{\beta}
\stackrel{d_2}{\longrightarrow}\oplus_{\beta}a^{(2)}_{\ap\beta}{\ca R}_{\beta}
\stackrel{d_1}{\longrightarrow}\oplus_{\beta}a^{(3)}_{\ap\beta}{\ca R}_{\beta}
={\ca R}_\ap\right]. \label{frc}
\ea
One of the results shown in \cite{ItNa} is that the $K$-group 
(Grothendick group)
 of bounded complex ${\ca S}_\ap$ has a support on the exceptional locus and 
 defines a basis of $K^c(X)$, where $K^c(X)$ is
  the $K$-group for complexes on the exceptional locus. Furthermore the 
  intersection between ${\ca R}_\ap$ and ${\ca S}_\beta$ is shown to be
$\la {\ca R}_\ap,{\ca S}_\beta \lb=\delta_{\ap\beta}$. From this and the 
complex (\ref{frc}) one obtains the intersection between ${\ca S}_\ap$ as
follows
\ba
\la {\ca S}_\ap,{\ca S}_\beta \lb=a^{(2)}_{\ap\beta}-a^{(1)}_{\ap\beta}
=a^{(1)}_{\beta\ap}-a^{(1)}_{\ap\beta}.
\ea
Then in string theory one can identify ${\ca S}_\ap$ as a fractional D0-brane 
$\rho_{\ap}$. 

We argue that this complex (\ref{frc}) just represents the tachyon 
condensation of D6-branes ${\ca R}_\ap
\oplus_{\beta}a^{(2)}_{\ap\beta}{\ca R}_{\beta}$ and anti D6-branes
${\ca R}_\ap
\oplus_{\beta}a^{(1)}_{\ap\beta}{\ca R}_{\beta}$. Indeed for the orbifold 
${\bf \Gamma}={\bf Z}_N$ this is the same as 
the previous result for the tachyon field $p=q=r=1$
in BSFT with ${\ca N}=2$ boundary supersymmetry. 
One can understand the interpretation of the arrows in the complex 
as tachyon fields in detail. Let us remember the explicit form of non-abelian
tachyon field (\ref{tm}). Then it is easy to see that the multiplication of
the tachyon fields $T_1=Z_1,\ T_2=Z_2,\ T_3=Z_3$ transform a D6-brane 
$\rho_{\ap}$ into three 
anti D6-branes $\rho_{\ap+a_1}\oplus\rho_{\ap+a_2}\oplus\rho_{\ap+a_3}$ and 
so on. This is equivalent to the arrows (the boundary operator 
$\stackrel{d_i}{\longrightarrow}$) defined before\footnote{When the elements 
of 
the non-abelian tachyon field are
 anti-holomorphic, we interpret these as arrows in the opposite direction.}. 
In other words, 
the wedge product of $B$ corresponds to the multiplication of the operator
$\sum_{i}{\bf \Gamma}_iT_{i}$. This explicitly shows its nilpotency.
We can also consider more general 
tachyon fields for any $p,q,r$. From the previous analysis we obtain
the following complex:
\ba
\sum_{\beta_i=0}^{p_i-1}{\ca S}_{\ap+a_1\beta_1+a_2\beta_2+a_3\beta_2}
: {\ca R}_\ap \to 
\oplus_{i}{\ca R}_{\ap+p_ia_i}
\to\oplus_{k\neq i,j}{\ca R}_{\ap+p_ia_i+p_ja_j}
\to{\ca R}_{\ap+p_1a_1+p_2a_2+p_3a_3}.
\ea
Note that this complex can be obtained from the linear combination of 
(\ref{frc}) with appropriate `brane-antibrane annihilation' 
(or more precisely up to quasi-isomorphism).

For non-abelian orbifolds we cannot show this 
correspondence explicitly since it is
 difficult to construct the appropriate boundary interaction.
However it is natural from the above results to believe that 
 any tachyon condensation which corresponds to the complex 
 (\ref{frc}) preserves ${\ca N}=2$ boundary supersymmetry. In this way
 the tachyon condensation in orbifold theory gives an interesting example
 which relates the string field theory restricted by ${\ca N}=2$ 
 boundary supersymmetry to mathematics on
 complex manifolds.

\section{Comment on ${\ca{N}}=4$ Boundary Supersymmetry}
\hspace{5mm}
\setcounter{equation}{0}
\hspace{5mm}
As we have seen, ${\ca{N}}=2$ supersymmetry in tachyonic boundary 
interactions
 enables a refined mathematical viewpoint. Thus it will also be interesting
 to consider ${\ca{N}}=4$ supersymmetry further. 
 The world-sheet ${\ca{N}}=4$
 supersymmetry constrains the target space ${\ca M}$ to be hyperkahler as is 
 well-known \cite{AlFr} and thus we assume that the dimension of ${\ca M}$
 is a multiple of four.
 Here we would like to discuss world-sheet with a 
 boundary which preserves ${\ca{N}}=4$ boundary supersymmetry. In particular
we discuss the Neumann boundary condition 
$\de_{1} \Phi^i|_{\de{\Sigma}}=0,\ \ 
(\psi_{L}^i-\psi_{R}^i)|_{\de{\Sigma}}=0\ $ since we are
interested in brane-antibrane systems wrapping on the whole manifold 
${\ca M}$ with no flux. 
The generalization of our results to other boundary conditions is 
obtained by the $SO(3)$ rotation (see \cite{OoOzYi}) directly.  
Since the appropriate ${\ca{N}}=4$ off-shell superspace is not known, 
 we work within 
 ${\ca{N}}=1$ superspace or by using its component expressions. 

The ${\ca{N}}=4$ non-linear sigma model is defined by the following action
 \cite{AlFr}:
\ba
I_{0}&=&\f{1}{4i}\int_{\Sigma} 
(d\sigma)^2(d\theta)^2 g_{ij}({\bf \Phi})\bar{D}
{\bf \Phi}^i
D{\bf \Phi}^j, \label{sm4}
\ea
where ${\bf \Phi}^{i}(\sigma)=\Phi^{i}(\sigma)+\theta\psi^{i}(\sigma)+\ddd$
denote the ${\ca{N}}=1$ superfields on the world-sheet $\Sigma$. 
Then this bulk action
is invariant under the ${\ca{N}}=4$ supersymmetry. Its restriction on the 
boundary (along $x^0$) of the world-sheet is given by
\ba
\delta\phi^i&=&i\ep\psi^{j}f^{(a)i}_{j}, \no
\delta\psi^i&=&-h^{(a)i}_j\ep(\de_{0}\phi^j)-i\Gamma^{i}_{kl}
f^{(a)l}_{j}\ep\psi^{j}
\psi^{k},\label{sh}
\ea
where $f^{(a)l}_{j}=(h^{(a)-1})^i_j, \ \ (a=1,2,3)$ denotes three independent
complex structures and this defines a hyperkahler 
structure. In other words these satisfy
\ba
f^{(a)i}_{j}f^{(b)j}_{k}+f^{(b)i}_{j}f^{(a)j}_{k}=
-2\delta^{ab}\delta^{i}_{k},\ \ \ 
g_{ij}f^{(a)i}_{k}f^{(a)j}_{l}=g_{kl},\ \ \
\nabla_{i}\ f^{(a)j}_{k}=0.
\ea
The first equation shows the correct algebra of 
2D ${\ca{N}}=4$ supersymmetry. The second and third are required by
the invariance of the action (\ref{sm4}) by the transformations (\ref{sh}).

The boundary interaction for tachyon fields is given by
\ba
I_{B}=\int_{\de\Sigma} (d\tau)(d\theta)({\bf \Gamma}^{A}D{\bf \Gamma}^{A}
+T^A({\bf\Phi})
{\bf \Gamma}^{A}).
\ea
Note that here we have employed ${\ca{N}}=1$ boundary superspace and thus
the boundary fermionic superfields ${\bf \Gamma}^{A}=\eta^{A}+\theta F^{A}$ 
are real. We would like for the combined action $I_{0}+I_{B}$ to be 
invariant 
under the 
${\ca{N}}=4$ boundary supersymmetry (\ref{sh}) and
\ba
\delta\eta^{A}&=&\ep F^{B}{\mathsf{f}}^{(a)A}_{B}, \no
\delta F^{A}&=&-i\ep\de_{\tau}\eta^{B}{\mathsf{h}}^{(a)A}_{B}.
\ea
This is possible if the following conditions are satisfied
\ba
{\mathsf{f}}^{(a)A}_{B}{\mathsf{h}}^{(a)B}_{C}&=&\delta^{A}_{C}, \label{s1} 
\\
{\mathsf{f}}^{(a)A}_{B}{\mathsf{f}}^{(b)B}_{C}+
{\mathsf{f}}^{(b)A}_{B}{\mathsf{f}}^{(a)B}_{C}&=&
-2\delta^{ab}\delta^{A}_{C}, \label{s2} \\
{\mathsf{f}}^{(a)A}_{B}&=&{\mathsf{h}}^{(a)B}_{A}, \label{s3} \\
\de_{i}{\mathsf{f}}^{(a)A}_{B}&=&0, \label{s4}\\
f^{(a)i}_{j}(\de_{i}T^{A})-(\de_{j}T^{B}){\mathsf{f}}^{(a)A}_{B}&=&0 
\label{s5}.
\ea
The equations (\ref{s1}) and (\ref{s2}) ensure the correct algebra of the 
boundary ${\ca{N}}=4$ supersymmetry. The others are needed for the 
invariance of the 
action $I_{0}+I_{B}$ under this supersymmetry. As can be seen from these
 requirements the three matrices ${\mathsf{f}}^{(a)A}_{B}\in O(4)
\ \ (a=1,2,3)$
  , which is constant due to eq.(\ref{s4}), define a 
  hyperkahler structure on the vector space $V$ along the boundary
  superfields ${\bf \Gamma}^A$. Thus the dimension of this vector space 
  should be a multiple of four again. Finally, the last equation (\ref{s5})
  constrains the tachyon fields so as to preserve the extended supersymmetry.
If one requires only ${\ca{N}}=2$ supersymmetry ($a=1$), then this means
that the tachyon fields are holomorphic and reproduce the result 
in \cite{Ho}. In the ${\ca{N}}=4$ case now considered, this gives a more 
strong restriction. For simplicity 
let us assume $dim(V)=dim({\ca M})=4$ and ${\ca M}$ is flat. 
By using $O(4)$ rotation we can set
 ${\mathsf{f}}^{(a)}=f^{(a)}$ and the explicit form can be given using
 the Pauli matrices as follows:
\ba
{\mathsf{f}}^{(1)}=(i\sigma_2)\otimes 1,\ \ 
{\mathsf{f}}^{(2)}=\sigma_1\otimes (i\sigma_2),\ \ 
{\mathsf{f}}^{(3)}=\sigma_3 \otimes(i\sigma_2).
\ea
Then it is easy to find the following linear solutions to eq.(\ref{s5}):
\ba
T^A(\Phi)=\sum_{i=1}^{4}M^{A}_{i} \Phi_{i},
\ea
where the matrix $M$ is any linear combination of both the identity and the 
following three matrices, which represent the quaternionic algebra:
\ba
{\mathsf{I}}=1\otimes (-i\sigma_2),\ \ 
{\mathsf{J}}=(-i\sigma_2)\otimes\sigma_3,\ \ 
{\mathsf{K}}=(-i\sigma_2)\otimes\sigma_1.
\ea
If we apply this result to the previous arguments on tachyon condensation
on two dimensional orbifolds, we find 
that these allowed configurations include the decay
of $\mbox{D}4-\ov{\mbox{D}4}$ into a fractional D0-brane which corresponds
to the following complex similar to (\ref{frc}):
\ba
{\ca S}_\ap: \ {\ca R}_\ap \stackrel{d_2}{\longrightarrow} 
\oplus_{\beta}a^{(1)}_{\ap\beta}{\ca R}_{\beta}
\stackrel{d_1}{\longrightarrow}{\ca R}_\ap.
\ea
For more general configurations of 
tachyon fields we have not
obtained any definite solutions of eq.(\ref{s5}).

If gauge fields on D-branes have a
  non-trivial configuration, then this space $V$ 
  is twisted and the boundary
  superfields ${\bf \Gamma}^A$ should be viewed as a special kind of 
  section (satisfying a restriction like eq.(\ref{s5})) 
  of a vector bundle on the hyperkahler manifold ${\ca M}$. It would 
  also be 
  interesting to construct ${\ca{N}}=4$ boundary interactions including
gauge fields and clarify the meaning of the `quaternionic' constraint
(\ref{s5}).

\section{Conclusions}
\hspace{5mm}
In this paper we have studied a mathematical aspect of 
tachyon condensation on brane-antibrane
systems in the (three dimensional) 
orbifold theories. We have applied to this system the boundary string
field theory with boundary ${\ca{N}}=2$ supersymmetry. As a result 
we have observed that
the essential part of the tachyon condensation can be understood as the
McKay correspondence. This example shows that
the ${\ca{N}}=2$ supersymmetry enables us to employ the 
language of homological algebra in algebraic geometry. One can regard this
as an example of the recent interpretation of D-branes as derived category. 
We have also discussed the boundary ${\ca{N}}=4$ 
supersymmetry and observed this symmetry leads to a certain quaternionic
constraint on tachyon fields.

\bigskip

\begin{center}
\noindent{\large \bf Acknowledgments}
\end{center}
I am very grateful to T. Eguchi and Y. Matsuo 
for valuable discussions and encouragements. 
I also thank 
T. Muto for explaining useful facts on orbifolds and thank S. Terashima,
M. Hamanaka and H. Kajiura for helpful discussions.
This work is supported by JSPS Research Fellowships for Young 
Scientists.

\end{document}